\documentclass[letters]{fcs}

\usepackage{placeins}
\usepackage{xcolor}\newcommand{\revise}[1]{#1}
\newcommand{\paperfivecommit}{\texttt{ac4966bdf5e5}}
\newcommand{\paperfiveurl}{\href{https://github.com/Gottesman-Software/lidmas_cpp/tree/ac4966bdf5e588eb9191388dd258cb2701b14a59/examples/paper_runs/paper_05/revision_evidence}{\texttt{Gottesman-Software/lidmas\_cpp: paper\_05/revision\_evidence}}}

\title{Auditing Syndrome-to-Decoder Mappings \\for Repetition, Surface, CSS, and Digitized-GKP Interfaces}
\shorttitle{Auditing Syndrome-to-Decoder Mappings}

\author[1,2,+]{Dennis~Delali~Kwesi~Wayo}
\author[3]{Chinonso~Onah}
\author[4]{Rodrigo~Alves~Dias}
\author[5]{Leonardo~Goliatt}
\author[2,+]{Sven~Groppe}
\address[1]{College of Computing, Georgia Institute of Technology, Atlanta, GA 30332, USA}
\address[2]{Faculty of Mathematics and Computer Science, TU Bergakademie Freiberg, Bernhard-von-Cotta-Stra\ss e 2, D-09599 Freiberg, Germany}
\address[3]{Department of Physics, RWTH Aachen University, Aachen, Germany}
\address[4]{Department of Physics, Federal University of Juiz de Fora, Juiz de Fora, 36036-900, Brazil}
\address[5]{Department of Computational and Applied Mechanics, Federal University of Juiz de Fora, Juiz de Fora, 36036-900, Brazil}
\corremail{dwayo3@gatech.edu; sven.groppe@informatik.tu-freiberg.de}
\fcssetup{
  doi = {10.1007/s11704-026-61541-6},
  article-id = {FCS-261541},
  first-page = {1},
  received = {2 September 2026},
  publish-month = {September},
  publish-year = {2026},
}

\begin{document}

\section{Introduction}

Quantum error correction ends with classical data. Before a decoder can act,
each measured bit needs an agreed meaning: which stabilizer it represents,
how a hardware register is ordered, and which data variable a correction index
names. This agreement is easy to dismiss as software plumbing. Yet a reversed register, permuted check row, or stale transpilation map can make a decoder solve the wrong problem even when the circuit and decoder each work as
designed. Such an interface failure is different from estimating a threshold
or demonstrating a fault-tolerant memory~\cite{fowler2012proofmatching,acharya2023suppressing}.

The danger is also concrete in surface-code experiments. Hardware count keys are
commonly printed from the highest classical bit to the lowest, while analysis
code often indexes arrays in the opposite direction. If that convention is
left implicit, a mathematically valid correction can be attached to the wrong
physical location. An aggregate localization score may not expose the mistake,
especially when source noise and decoder degeneracy already spread corrections
over several sites.

This LETTER asks how the circuit-to-decoder agreement can be tested directly.
It uses a shared measurement-bit $\rightarrow$ syndrome-row $\rightarrow$
decoder-variable contract in four small, distinct branches: a
five-data-qubit repetition code; the $Z$-check layer of a \(5\times5\) planar
surface-check lattice; the seven-qubit Steane code as a compact
Calderbank--Shor--Steane (CSS) benchmark
\cite{calderbank1996good,steane1996error}; and a digitized Gaussian
displacement/readout proxy whose binary outer checks share the surface
incidence matrix. \revise{Here ``surface'' denotes this
surface-code-inspired \(Z\)-check interface only; the one-check-type
construction is not used to establish an encoded logical qubit or a code
distance.} These branches are not combined into a decoder benchmark.
They provide different ways to challenge the same interface contract. LiDMaS+
serves as the narrow request adapter that serializes ordered syndrome events, code
metadata, and decoder responses; its presence does not establish decoding
accuracy. \revise{This work is narrower than the general LiDMaS+
hardware-to-decoder architecture introduced previously
\cite{wayo2026unified}: the contribution here is an IBM-oriented
mapping-validation audit, with negative controls, for fixed small interfaces.}

The conformance test starts from known single-data-qubit $X$ injections rather
than from noisy hardware outcomes. \revise{A support-list/matrix builder,
separate from the circuit generator, supplies the expected syndrome for every
one of the 40 data positions. The only shared ingredients are the published
check supports and register convention.} A separate binary interpreter
\revise{that contains no calls to the decoder, mapping, or syndrome-extraction
routines under test} propagates the noiseless gate list, reconstructs the
measured classical register, and requires the extracted 16-bit vector to equal
the corresponding column of $H_{\mathrm{surf},Z}$. Reversing the 16
syndrome bits provides a targeted negative control. The audit is required to
recognize the intended map and reject the convention error it was designed to
find. Technical definitions, the full matrix, audit tables, and reproduction
commands are provided in Online Resource 1.

\section{Method}

For a binary check matrix $H$ and measured syndrome $s$, the decoder selects
the exact binary minimum-weight (EMW) solution
\begin{equation}
 \hat e=\arg\min_{e\in\{0,1\}^{n}}\{|e|:He=s\pmod 2\},
 \label{eq:emw}
\end{equation}
with lexicographic ordering of sorted data indices as the deterministic tie
break. This optimization is EMW, not minimum-weight perfect matching, because
the general matrices are not converted into matching graphs. Exploratory
union-find-like growth and hard-decision bit-flip policies are retained only in Online Resource 1, where their residuals are reported both before and after an explicitly
separated EMW closure step \cite{unionfind2017,duclos2010fast}.

The repetition, surface, and compact CSS circuits were executed once on
\texttt{ibm\_fez}, with 4096 shots per clean or injected stream. Their locally
sampled controls used an explicit 2\% independent measurement-bit flip rate,
zero background data-error rate, and seed 20260705; they are reported as
\emph{noisy local controls}, not ideal simulators. By contrast, the deterministic audit has no noise and no sampling. IBM Runtime job identifiers, optimization level,
register order, available software versions, and the limitation that the
original export did not retain calibration snapshots or exact ISA layouts are
given in Online Resource 1.

\begin{quote}
\footnotesize
\textit{Deterministic audit pseudocode.} For each target
\(q\in\{0,\ldots,39\}\): (1) build \(C_q\) with the injected \(X_q\)
and explicit measurements; (2) propagate the noiseless instruction list with an
independent binary interpreter; (3) reconstruct the high-to-low count key,
reverse it once, and slice \(s_q=(c_0,\ldots,c_{15})\) and
\(d_q=(c_{16},\ldots,c_{55})\); and (4) accept the target only if
\(s_q=H_{\mathrm{surf},Z}[:,q]\), \(d_q=e_q\), and
\(\mathrm{rev}(s_q)\neq H_{\mathrm{surf},Z}[:,q]\).
\end{quote}

The fourth branch is a classical Gaussian proxy for digitized
Gottesman--Kitaev--Preskill (GKP) readout. It does not simulate a finite-energy
grid state \cite{gkp2001,fukui2018highthreshold}. Its baseline uses three
rounds, displacement standard deviation $0.015\sqrt\pi$ per round, squeezed
quadrature variance $e^{-4}$ ($r=2$), half-cell jump probability $10^{-3}$,
measurement-flip probability $0.01$, injection $0.56\sqrt\pi$, decision
half-width $0.25\sqrt\pi$, 4096 shots, and seed 20260706. Figure~\ref{fig:audit}
varies one parameter at a time. \revise{This branch stresses the same mapping
contract from the analog side: continuous displacement/readout records are
thresholded into an ordered binary check vector before the decoder sees them,
so a binning, ordering, or register-convention error would violate the same
interface semantics tested in the qubit branches.}

\begin{figure*}[t]
 \centering
 \begin{minipage}[t]{0.39\textwidth}
 \centering
 \begin{tikzpicture}[x=0.075cm,y=0.55cm,font=\scriptsize]
   \foreach \label/\value/\yy in
     {Syndrome map/40/2,Permutation detected/40/1,Data round-trip/40/0}{
       \fill[fcsblue!78] (0,\yy-0.25) rectangle (\value,\yy+0.25);
       \node[anchor=east,align=right] at (-0.5,\yy) {\label};
       \node[anchor=west] at (\value+0.4,\yy) {\value/40};
     }
   \draw[->] (0,-0.55)--(42,-0.55) node[right]{targets};
   \foreach \xx in {0,10,20,30,40}
     \draw (\xx,-0.48)--(\xx,-0.62) node[below]{\xx};
 \end{tikzpicture}\par\smallskip
 {\small (a) Deterministic mapping audit}
 \end{minipage}\hfill
 \begin{minipage}[t]{0.58\textwidth}
 \centering
 \begin{tikzpicture}[x=7.1cm,y=0.43cm,font=\scriptsize]
   \draw[->] (0,-0.65)--(0.82,-0.65) node[right]{localization};
   \foreach \xx in {0,0.2,0.4,0.6,0.8}
     \draw (\xx,-0.58)--(\xx,-0.72) node[below]{\xx};
   \foreach \name/\elo/\ehi/\rlo/\rhi/\yy in {
     Baseline/.345/.498/.418/.610/5,
     {$p_m=0$}/.406/.583/.438/.633/4,
     {$p_m=.03$}/.256/.370/.369/.563/3,
     {$w=.20\sqrt\pi$}/.458/.526/.633/.765/2,
     {$w=.30\sqrt\pi$}/.160/.329/.205/.409/1,
     {$r=1.5$}/.328/.468/.399/.585/0}{
       \node[anchor=east] at (-0.02,\yy) {\name};
       \draw[fcsblue,line width=1.5pt] (\elo,\yy+0.09)--(\ehi,\yy+0.09);
       \draw[fcsblue] (\elo,\yy-0.04)--(\elo,\yy+0.22)
                         (\ehi,\yy-0.04)--(\ehi,\yy+0.22);
       \draw[fcsorange,line width=1.5pt] (\rlo,\yy-0.09)--(\rhi,\yy-0.09);
       \draw[fcsorange] (\rlo,\yy-0.22)--(\rlo,\yy+0.04)
                         (\rhi,\yy-0.22)--(\rhi,\yy+0.04);
     }
   \draw[fcsblue,line width=1.5pt] (0.07,5.65)--(0.12,5.65)
     node[right,black]{exact};
   \draw[fcsorange,line width=1.5pt] (0.33,5.65)--(0.38,5.65)
     node[right,black]{target recall};
 \end{tikzpicture}\par\smallskip
 {\small (b) Digitized-GKP one-factor sensitivity}
 \end{minipage}
 \caption{Independent validation. (a) All noiseless single-$X$ circuits produce
 the required column of $H_{\mathrm{surf},Z}$; reversing the syndrome register
 is detected for every target, and measured data bits round-trip correctly.
 (b) Exact and target-containing localization ranges of the digitized-GKP proxy
 under one-factor variations. These parameter-dependent values are interface
 diagnostics, not physical GKP error-correction performance.}
 \label{fig:audit}
\end{figure*}

\section{Results}

The deterministic audit passed 40/40 syndrome mappings and 40/40 data-bit
round trips; the reversed-bit control failed the expected-column equality for
40/40 targets (Fig.~\ref{fig:audit}a). For the eight hardware surface streams,
the injected data bit was independently the most frequently activated data
measurement in every case. Its activation rate was 0.731--0.918, whereas the
mean non-target rate was 0.041--0.082. Together, these checks establish the register identity and noiseless circuit
semantics. They do not remove noise from the single hardware job.
\revise{The 40/40 result is limited to the stated noiseless single-\(X\),
single-round \(Z\)-check construction and its classical-register convention; it
does not validate repeated-round detector construction, complementary
check-type mappings, logical-observable mappings, or arbitrary faults.}

\begin{table}[t]
\centering
\caption{EMW localization over injected streams. ``Contains'' is target recall,
not logical success. FP is mean non-target correction support per shot.}
\label{tab:results}
\fontsize{7.2pt}{8.2pt}\selectfont
\setlength{\tabcolsep}{1.8pt}
\begin{tabular}{@{}lccc@{}}
\toprule
Source & Exact & Contains & FP support\\
\midrule
Repetition, IBM & .821--.868 & .876--.923 & .115--.210\\
Repetition, 2\% local & .910--.927 & .933--.961 & .084--.110\\
Surface, IBM & .003--.108 & .279--.642 & 1.984--3.134\\
Surface, 2\% local & .715--.728 & .865--.953 & .355--.407\\
Compact CSS, IBM & .843--.858 & .843--.858 & .067--.152\\
Compact CSS, 2\% local & .938--.944 & .938--.944 & .036--.062\\
Digitized-GKP proxy & .350--.495 & .417--.608 & .289--.943\\
\bottomrule
\end{tabular}
\end{table}

Table~\ref{tab:results} separates source quality from mapping validity. The
surface hardware stream has only 0.112--0.297 mean support precision and
1.984--3.134 false-positive correction sites per shot, despite the target
appearing in some correction sets. Target-containing localization is instead
reported as recall of an injected position, not as successful error
correction. Logical equivalence is not evaluated because this experiment
executes only one check type and does not contain the complete stabilizer and
logical-observable record. \revise{For singleton injected truths, support
precision is mathematically identical to the Jaccard score reported in Online Resource 1:
when \(q\in\hat e\) both equal \(1/|\hat e|\), and otherwise both are zero.}

The localization ranges depend strongly on the GKP-proxy parameters. Removing measurement
flips raises exact localization to 0.406--0.583, while $p_m=0.03$ lowers it to
0.256--0.370; changing the decision half-width from $0.20\sqrt\pi$ to
$0.30\sqrt\pi$ changes exact ranges from 0.458--0.526 to 0.160--0.329. The
proxy shows only that analog-derived records can obey the
same ordered binary request contract. \revise{Online Resource 1 distinguishes the
archived baseline stream used for Table~\ref{tab:results} from the separately
regenerated common-random-number sensitivity baseline in
Fig.~\ref{fig:audit}b; their small numerical differences are Monte Carlo stream
differences, not different physical parameters.}

\section{Conformance and reproducibility}

\revise{The mapping audit and sampled localization study answer different
questions. The audit checks whether the Qiskit gate/register semantics
reproduce the prescribed matrix column: a support-matrix builder separate from
the circuit generator supplies truth, while a binary interpreter with no calls
to the mapping or decoder routines propagates the instruction list and
reconstructs count-key order. The 40/40 pass validates only this
single-\(X\), single-round \(Z\)-check circuit-to-register map; the 40/40
reversed-register failures show sensitivity to the targeted convention error.}

\revise{Native heuristic failures are reported before closure in Online Resource 1:
IBM repetition/surface/CSS rates are 0.058/0.416/0, noisy-local rates are
0.034/0.068/0, and the digitized-GKP proxy rate is 0.162. Each invoked closure
adds a weight-two EMW correction in these records, while EMW itself has no
closure. Exact localization remains representative-dependent under EMW's
lexicographic tie break, so recall, false-positive support, and precision are
localization metrics, not logical success. The three IBM families are fixed
streams from jobs \texttt{d952dssql68s73ca4ojg},
\texttt{d953c7mvtlqs73fuuk90}, and \texttt{d9577ptgc6cc73ffs8hg}; they
preserve counts, optimization level, register order, and software versions, but
not ISA layouts, depth/gate counts, calibration snapshots, or verified
execution timestamps. Missing metadata are not reconstructed; \paperfiveurl{}
fixes the code and supplies the audit and sensitivity files.}

\section{Discussion and conclusion}

The experiment establishes a narrow result: under noiseless propagation, the
implemented measurement map reproduces every surface-check column and detects
a targeted register reversal. The sampled local and hardware streams answer
a different question--how often a noisy record leads the fixed EMW
representative back to an injected site. Those localization rates are not
decoder benchmarks, threshold estimates, or logical-QEC success probabilities.

The hardware evidence has narrower scope. Wilson intervals in Online Resource 1
describe binomial uncertainty within one fixed stream, but each IBM family came
from a single Runtime job. Calibration drift, compilation variability, and
between-day effects remain unmeasured. The archived surface stream shows
broad syndrome activation; the available data do not identify its cause.
Resolving that cause would require repeated jobs with preserved layouts,
circuit metrics, and calibration snapshots.

In a decoder benchmark, the first evidentiary step is producer-decoder
agreement on the meaning of every measured bit. Known single-site injections test the positive
path, an independently assembled syndrome supplies ground truth, and a
permutation negative control shows that the audit can detect the convention
error of interest. Native residuals reported before any closure step keep
failures visible. Otherwise, a guaranteed repair can make an exploratory policy appear
exact and conceal where the original rule failed.

The same contract can be extended to repeated-round detectors, both check
types, and logical-observable evaluation, provided that every layer publishes
its bit order and parity convention. The present data establish
single-check-type localization under fixed streams. A logical-performance study
would additionally need independent decoder implementations, repeated hardware
runs, retained ISA layouts and calibrations, and corrections scored modulo the
full stabilizer group.

Across the four branches, the result is stable. A bit reversal fails the
ground-truth audit; source noise lowers localization without masquerading as a
mapping failure; and a closure step cannot erase the record of native
residuals. These checks do not turn the experiment into a logical-QEC
benchmark. They establish a prerequisite for one: a decoder output is
meaningful only when the declared map assigning meanings to measured bits remains correct throughout.

\section*{Acknowledgments}
The authors thank users of the decoder-replay tooling for feedback on interface
diagnostics.

\section*{Competing interests}
The authors declare no competing interests. No external funding was received.

\section*{Online Resource 1}
Online Resource 1 contains the full surface construction and matrix, enlarged
circuit renderings, IBM job and software metadata, mapping and negative-control
rows, decoder pre-closure audits, support metrics, GKP parameters and sensitivity,
and reproduction commands. The audit and sensitivity files are also available
from the permanent revision-evidence link in the Data and code availability
statement.

\section*{Data and code availability}
The workflow source, reproduction commands, decoder policies, analysis, and
plotting scripts are fixed at \paperfiveurl{} (commit \paperfivecommit).
Online Resource 1 reports the audit results and IBM job identifiers used in this study.
The fixed raw count exports \revise{and generated decoder-response streams}
are retained by the corresponding author \revise{and can be supplied upon
reasonable request, subject to platform terms}; they are not included in the
public repository, and credentials are excluded.

\fontsize{8.5pt}{10pt}\selectfont

\end{document}